# Selective-injection GaN Heterojunction Bipolar Transistors with 275 kA/cm² Current Density

Zhanbo Xia[1, *], Chandan Joishi[1, *], Shahadat H. Sohel[1], Andy Xie[2], Edward Beam[2], Yu Cao[2], Siddharth Rajan[1, 3]

[1]Department of Electrical and Computer Engineering, The Ohio State University, Columbus, OH 43210 USA

[2]Qorvo, Inc, Richardson, TX 75081, USA

[3]Department of Materials Science and Engineering, The Ohio State University, Columbus, OH 43210 USA

Email: joishi.1@osu.edu rajan.21@osu.edu

**Abstract**— We design and demonstrate selective injection GaN heterojunction bipolar transistors that utilize a patterned base for selective injection of electrons from the emitter. The design maneuvers minority carrier injection through a thin p-GaN base region, while the majority carrier holes for base current are injected from thick p-GaN regions adjacent to the thin p-GaN base. The design is realized using a regrowth emitter approach with $SiO_2$ as a spacer between the emitter layer and the thick p-GaN base contact regions. The fabricated device demonstrated state-of-art output current density ($I_{C, max}$) ~275 kA/cm² with a current gain (β) of 9, and 17 for the planar HBT design ($I_{C, max}$ =150 kA/cm²). The reported results highlight the potential of the selective injection design to overcome the persistent GaN HBT design tradeoff between base resistance and current gain, paving the way for next-generation radio frequency and mm-Wave applications.

## I. Introduction

THE Gallium Nitride (GaN) high electron mobility transistors (HEMTs) in its various forms, is currently the device of choice for numerous power and radio-frequency electronics [1-4]. These devices leverage the high breakdown field and superior transport properties of GaN to achieve performance metrics that significantly surpass traditional silicon technology [5]. However, the premature breakdown (~1 MV/cm) in these devices due to a non-uniform gate-drain electric field, positioning of peak fields at the semiconductor surface, and charge trapping from surface states impact their dynamic characteristics and long-term reliability [6], while the inherent trade-offs between breakdown and gain resulting from a field-plate design impede their high-power handling potential at elevated frequencies [7].

GaN heterojunction bipolar transistors (HBTs), on the other hand, offer a path to overcome these challenges. The HBT is least impacted by surface states since the peak field occurs in the material bulk distant from the semiconductor surface [8]. As such, GaN HBTs can operate close to the theoretical breakdown limit of the material (~3.3 MV/cm).

Moreover, the HBT's vertical geometry confers higher power handling capabilities (due to volumetric injection of current) and better linearity due to a linear relationship between collector current and current gain [9], while also offering the highly sought-after normally-OFF operation in its default configuration. In addition, the base channel in an HBT can be engineered along the direction of carrier transport through heterojunction band engineering. Furthermore, the vertical topology of HBTs could enable easier integration solutions of III-Nitride devices on CMOS platforms for mixed signal applications while avoiding strain-related failures caused by thermal mismatch inherent to GaN-based lateral devices [10].

Despite their advantages, achieving high-performance n-p-n GaN-based HBTs has been challenging due to the low conductivity of the p-type base layers. While GaN HBTs with p-InGaN as the base benefit from enhanced base conductivity, they suffer from significant base-collector leakage due to lattice mismatch [11]. In contrast, p-GaN base offer low base-collector reverse leakage [12-14] but are constrained by the high sheet resistivity of the p-GaN layer. Thick p-GaN layers, intended to improve base conductivity, result in low current gain and high base delay. In contrast, thinner base regions result in current crowding, which limits current density and leads to high emitter-base capacitance charging delays [15]. To date, GaN HBTs have not reached their anticipated performance potential, with output current densities < 30 kA/cm$^2$ and moderate current gain [11, 14, 16-24], pulsed current densities < 150 kA/cm$^2$ [20], unity current gain cutoff frequencies ($f_T$) < 8 GHz, and maximum oscillation frequencies ($f_{max}$) < 2 GHz [16].

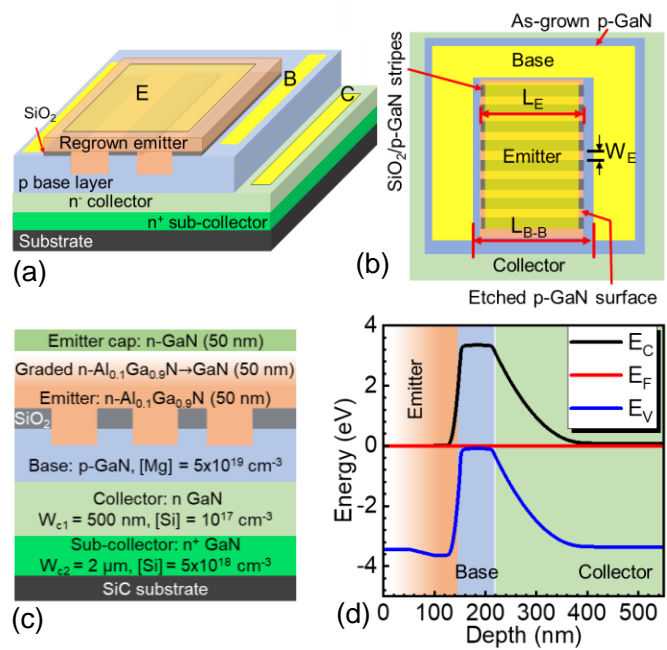

Fig. 1. (a) Three dimensional schematic, (b) top view, (c) two-dimensional schematic, and (d) energy band diagram of the proposed GaN SI-HBT.

A novel approach to enhancing HBT performance by addressing the trade-off between base resistance and current gain involves designing the base layer to facilitate selective injection of minority carriers from the emitter to the collector [25]. In this manuscript, we design and demonstrate a selective injection GaN HBT (SI-HBT) using a patterned emitter regrowth approach. The design features interdigitated thick and thin p-GaN base stripes with SiO$_2$ serving as a blocking layer to prevent electron injection from the emitter into the thick base layers. The electrons are only injected in the thin p-GaN regions to reduce base transit time and improve current gain, while base ohmic contacts are fabricated on the thick p-GaN layers to lower the base contact resistance. The holes for the base current are injected through the thick p-GaN layers and recombine with electrons in the thin regions. This design could further alleviate emitter crowding, particularly as the thin p-GaN stripes are reduced to sub-micron dimensions. Using the emitter regrowth approach, we achieved an output current density of 275 kA/cm$^2$ for the alternating thick/thin p-GaN stripe design, with current gain ($\beta$) < 10. The current gain obtained is the highest ever reported for a III-Nitride HBT.

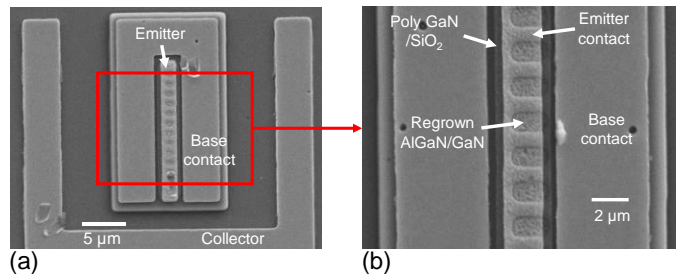

**Fig. 2.** Top view SEM images of the fabricated SI-HBT.

## II. DEVICE GROWTH AND FABRICATION DETAILS

The two and three-dimensional schematics of the fabricated SI-HBT and the corresponding energy band diagram are shown in Figure 1. Metal organic chemical vapor deposition (MOCVD) was used to grow the HBT epitaxial stack on a semi-insulating 4H-SiC substrate. The growth was performed at Qorvo. The epitaxial structure comprises of 1 μm n-GaN as the sub-collector with Si doping density, [Si] = 5×10$^{18}$ cm$^{-3}$, 500 nm n-GaN collector with [Si] = 10$^{17}$ cm$^{-3}$, and 150 nm p-GaN with [Mg] = 5×10$^{19}$ cm$^{-3}$. To define the emitter, 300 nm SiO$_2$ was deposited using plasma enhanced chemical vapor deposition (PECVD) on the p-GaN surface to function as a hard mask for emitter regrowth. The interdigitated emitter stripes ($L_E \times W_E$ in dimension separated from each other by $W_E$) were then defined using photolithography and SiO$_2$ along with 85 nm p-GaN was etched from the patterned windows. The etching of SiO$_2$ and GaN was carried out using CF$_4$/O$_2$ and BCl$_3$/Cl$_2$ etch recipes (respectively) in an Inductively coupled plasma reactive ion etching (ICP-RIE) chamber. Thereafter, the sample was treated with 25% TMAH at 80 °C for 10 mins and room temperature piranha for 10 mins to mitigate plasma etch-damage on the p-GaN sidewalls and the p-GaN surface [26]. The emitter layer was then regrown using plasma assisted molecular beam epitaxy (PAMBE). The emitter consists of 50 nm Al$_{0.1}$Ga$_{0.9}$N ([Si] = 5×10$^{18}$ cm$^{-3}$), 50 nm Al$_{0.1}$Ga$_{0.9}$N graded to GaN ([Si] = 10$^{19}$ cm$^{-3}$), and 50 nm GaN ([Si] = 10$^{19}$ cm$^{-3}$) as the emitter contact layer.

Following emitter regrowth, the collector is first defined by etching the regrown emitter layer, SiO$_2$, and ~750 nm GaN underneath the SiO$_2$ from the extrinsic emitter-base areas using BCl$_3$/Cl$_2$ based dry etch for GaN and AlGaN, and buffered HF (10:1) for SiO$_2$. Thereafter, the devices were annealed in a rapid thermal annealing (RTA) chamber at 850 °C for 12 mins in N$_2$ ambient. This step was seen to improve the current-voltage (I-V) characteristics of the base-collector diode as discussed later. To form the base contact, Pd/Ni/Au was deposited as the ohmic metal stack followed by rapid thermal annealing at 400 °C for 1 min in N$_2$. Finally, Al/Au was deposited as the collector and emitter metal contacts. Scanning electron microscopy images (SEM) of the SI-HBT with an emitter length ($L_E$) of 2 μm and width

($W_E$) of 1 µm is shown in Fig. 2.

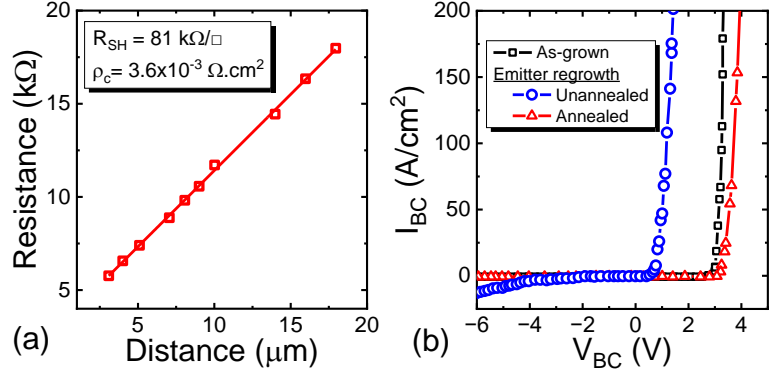

**Fig. 3.** (a) Transfer length method (TLM) measurements of the base contacts, (b) base-collector diode I-V characteristics of the device.

III. RESULTS AND DISCUSSIONS

The electrical characteristics of the SI-HBT were measured using a Keysight B1500 semiconductor device parameter analyzer. Transfer length method (TLM) measurements of the base ohmic contact on the thick p-GaN base layer (150 nm) is shown in Fig. 3(a). A base sheet resistance of 81 kΩ/□, with specific sheet resistivity of $3.6\times10^{-3}$ Ω.cm² was extracted from the TLM measurements. Two terminal I-V characteristics measured on the SI-HBT base-collector diode before and after the emitter regrowth is plotted in Fig. 3(b). As seen from the plot, the as-grown base-collector junction shows excellent characteristics with a turn-on voltage ~ 3 V, and $I_{ON}/I_{OFF}$ ~ $10^{10}$ (not shown here). Significant degradation was observed in the base-collector I-V post emitter regrowth with turn-on voltage ~ 1 V and high reverse bias leakage. We hypothesize that the degradation is due to unintentional impurities at the MBE regrowth interface which serves as a leakage pathway [27], and/or p-GaN deactivation by hydrogen present in the PECVD $SiO_2$ mask during the regrowth. Post regrowth annealing at 850 °C for 12 mins was seen to recover the P-N junction behavior, with similar turn-on and $I_{ON}/I_{OFF}$ ratio compared to the as-grown diode. A slight increment in the ON-resistance was also observed after the emitter regrowth.

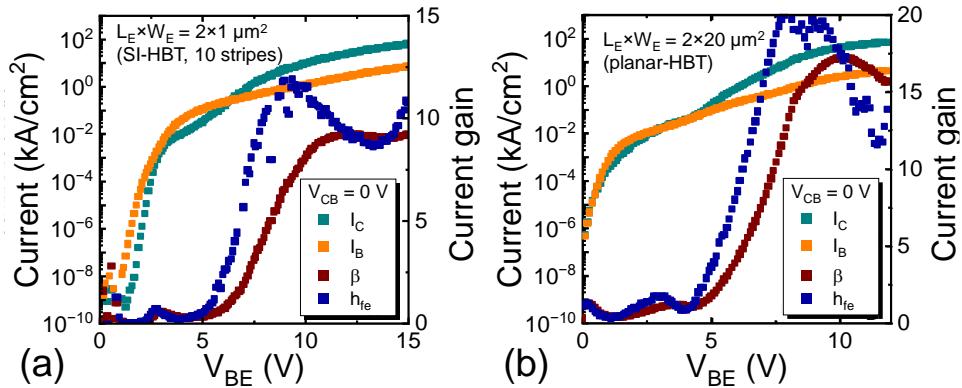

**Fig. 4.** Gummel plot of the (a) SI-HBT ($L_E \times W_E$ = 2×1 µm², 10 stripes), and (b) planar emitter HBT (emitter area = 2×20 µm²)

The Gummel plot of the SI-HBT ($L_E \times W_E$ = 2×1 µm², 10 stripes, total emitter mesa area = 2×20 µm²), measured at a collector-base voltage ($V_{CB}$) of 0 V, is shown in Fig. 4(a). For comparison, a planar HBT with similar dimensions as an SI-HBT without the emitter stripes was also fabricated alongside SI-HBT. From the plot, the common emitter current gain (β) peaked at ~9 for the SI-HBT and ~17 for the planar HBTs at high collector current densities. Besides, peak small signal AC gain ($h_{fe}$) of ~12 and ~20 was observed for the respective devices. The current gain is sensitive to the p-GaN etch recipe, and the low current gain is attributed to p-GaN etch-induced damage, recombination at the non-ideal emitter-base regrowth interface and the etched sidewalls [28], and Webster effects at these high current densities [25, 29]. Further optimization of the regrowth interface and the hole sheet density is needed to improve the current gain of the devices. The output characteristics ($I_C$-$V_{CE}$) of the SI-HBT and the planar emitter HBT are shown in Figs. 5(a) and (b). A peak collector current density ($I_C$) of 275 kA/cm$^2$, normalized to the regrown emitter area, was obtained for the SI-HBT, while the planar emitter HBT showed a peak current density of 150 kA/cm$^2$. The difference in the maximum current density indicates the improvement in emitter current crowding by the selective-injection design. The high $V_{CE, offset}$, i.e., minimum collector bias needed to obtain a positive collector current, in the output characteristics is attributed to voltage drop in the lateral emitter-base access region ($L_{B-B} - L_E \sim 1$ μm) [25, 30], while the high knee voltage observed is due to the unoptimized base contact resistance. Nevertheless, this is the highest current density ever reported for a III-Nitride HBT.

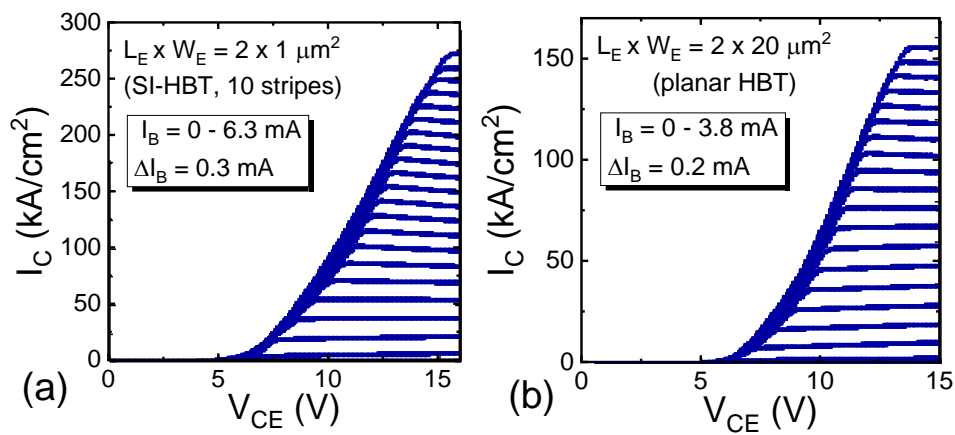

**Fig. 5.** Output characteristics of the (a) SI-HBT and (b) planar emitter HBT.

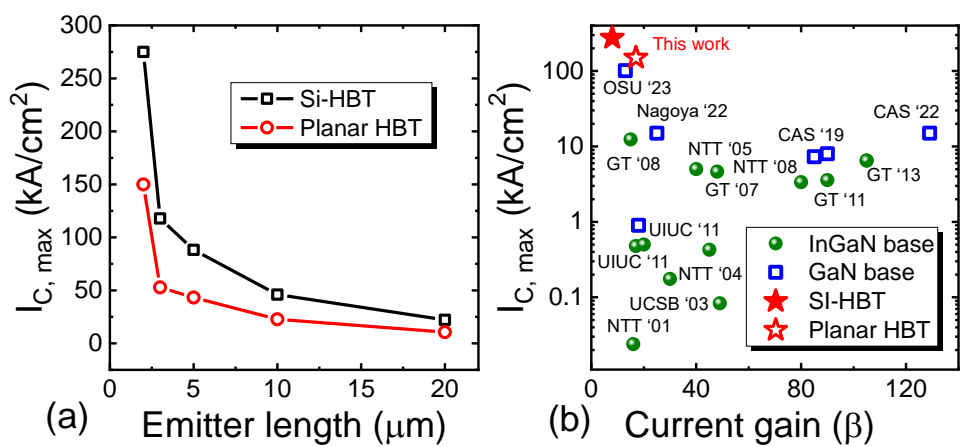

**Fig. 6.** (a) $I_{C, max}$ as a function of the emitter length and (b) benchmarking of $I_{C, max}$ with reported literature for III-Nitride HBTs.

The impact of emitter length ($L_E$) on $I_C$ for the same width ($W_E$ = 1 μm) is plotted in Fig. 6(a). $I_C$ was observed to decrease with increase in the emitter length. This highlights the impact of current crowding to device characteristics, i.e. while the current crowding was suppressed by moving from planar to SI-HBT architecture, complete mitigation of the crowding effects necessitates the emitter dimensions to be scaled to sub-micron scale.

Fig 6(b) shows a benchmark plot comparing $I_C$ vs $\beta$ obtained in this work with previously reported data on GaN HBTs [3, 4, 12, 14, 16, 17, 19-25, 31]. The DC output current densities reported across all III-Nitride HBT device designs till date are below 30 kA/cm$^2$ with recent report of $I_C$ = 101 kA/cm$^2$ by our group using a base regrowth approach to selective injection [25]. The present work with $I_C$ = 275 kA/cm$^2$, as such, is the highest output current density reported till date. While the observed current gain is low for reasons explained in the manuscript, improved base epilayer designs using polarization induced two- and three- dimensional hole gases in the thin and thick p-base stripes to further improve base conductivity and base contact resistance, improvement in the p-GaN etch process as well as the regrowth interface, and submicron stripe dimensions are expected to further improve the performance metrics of the device.

## IV. Conclusion

In conclusion, we demonstrated GaN HBTs that utilize selective minority carrier injection through periodically patterned thick and thin p-GaN base stripes to significantly enhance device performance. Using the patterned emitter regrowth approach with SiO$_2$ as the hard mask during the regrowth as well as the blocking layer for minority carrier injection from the emitter to the extrinsic base, we demonstrated state-of-art output current densities of 275 kA/cm$^2$. Although the work underscores the need to reduce base resistance by enhancing hole density as well as improve the current gain, the achieved current density is highly encouraging and highlights the importance of continued research on III-Nitride HBTs to enable new functionalities in III-Nitride high-power and RF device design.